\documentclass[reprint,showpacs,preprintnumbers,amsmath,amssymb, citeautoscript,aip,superscriptaddress]{revtex4-1}
\usepackage{graphicx,wasysym}
\usepackage{dcolumn}
\usepackage{bm}
\setcitestyle{super}

\usepackage{tabulary}
\usepackage{hyperref}

\newcommand{\kB}{k_\mathrm{B}}

\newcommand{\ve}[1]{\mathbf{#1}}
\usepackage{bbold}

\begin{document}

\preprint{1}

\title{Photoinduced phase separation in the lead halides is a polaronic effect}
\author{David T. Limmer}
 \email{dlimmer@berkeley.edu}
 \affiliation{Department of Chemistry, University of California, Berkeley, California}
\affiliation{Kavli Energy NanoScience Institute, Berkeley, California}
\affiliation{Materials Science Division, Lawrence Berkeley National Laboratory, Berkeley, California}
\affiliation{Chemical Science Division, Lawrence Berkeley National Laboratory, Berkeley, California}
\author{Naomi S. Ginsberg}
 \affiliation{Department of Chemistry, University of California, Berkeley, California}
 \affiliation{Department of Physics, University of California, Berkeley, California}
\affiliation{Kavli Energy NanoScience Institute, Berkeley, California}
\affiliation{STROBE NSF Science \& Technology Center, University of California, Berkeley, California}
\affiliation{Materials Science Division, Lawrence Berkeley National Laboratory, Berkeley, California}
\affiliation{Molecular Biophysics and Integrated Bioimaging Division, Lawrence Berkeley National Laboratory, Berkeley, California}

\date{\today}
\begin{abstract}
We present a perspective on recent observations of the photoinduced phase separation of halides in multi-component lead-halide perovskites. The spontaneous phase separation of an initial homogeneous solid solution under steady-state illumination conditions is found experimentally to be reversible, stochastic, weakly dependent on morphology, yet strongly dependent on composition and thermodynamic state. Regions enriched in a specific halide species that form upon phase separation are self-limiting in size, pinned to specific compositions, and grow in number in proportion to the steady-state carrier concentration until saturation. These empirical observations of robustness rule out explanations based on specific defect structures and point to the local modulation of an existing miscibility phase transition in the presence of excess charge carriers. A model for rationalizing existing observations based on the coupling between composition, strain and charge density fluctuations through the formation of polarons is reviewed. 
\end{abstract}
\maketitle

Lead halide perovskites have emerged as a promising class of semiconducting materials for photovoltaic and light emitting devices.\cite{kovalenko2017properties, beal2016cesium, ha2014synthesis, mcmeekin2016mixed} They are solution processable, defect tolerant, and offer a large library of potential photoactive compositions.\cite{de2015impact,mcmeekin2016mixed,stranks2013electron,burschka2013j,nie2015high}  The chemical formula of the perovskite lattice is ABX$_3$, where the B site cation is lead, the A site cation can be inorganic, like Cs, or organic, like methylammonium (MA) or formamidinium (FA), and the X site anion is a halide, typically Cl, Br or I. Both single crystals and thin films have been synthesized containing mixtures of X site halides.  Solid solutions of different halide anions are particularly attractive as their intrinsic band gaps are monotonic functions of composition, increasing across the visible spectrum from 1.5 eV for pure I to 2.9 eV for pure Cl.\cite{noh2013chemical,sadhanala2015blue} However, upon continued light exposure, initially homogeneous solid solutions of mixed halides undergo spontaneous phase separation.\cite{hoke2015reversible} During photoinduced phase separation, low bandgap states form and act as traps with enhanced low energy luminescence. This behavior is detrimental to device performance, as the low energy trap states degrade power conversion efficiencies.\cite{samu2017victim} Observations of photoinduced phase separation have been made for broad ranges of composition and morphology, suggesting that this behavior is general and intrinsic to this class of material.\cite{slotcavage2016light,brennan2017light} As a consequence, it has garnered significant attention for both practical and fundamental reasons.  This perspective aims at summarizing existing observations of this phenomena and at reviewing a specific framework that we have developed within which these observations can be understood.

The existence of photoinduced phase separation reflects an unusual interplay between the electronic and lattice degrees of freedom in the lead halide perovskites that is distinct from traditional inorganic materials and organic semiconductors. Unlike traditional inorganic semiconductors, the lead halide octahedral crystal structure is held together by predominately isotropic ionic bonds, \cite{egger2016hybrid} which results in a lattice that is anharmonic,\cite{brivio2015lattice, marronnier2018anharmonicity,yaffe2017local,beecher2016direct} mechanically soft,\cite{rakita2015mechanical} accommodate high defect concentrations\cite{kim2014role} and correspondingly large ionic diffusivities.\cite{lai2018intrinsic, kerner2017ionic, peng2018quantification} Unlike organic semiconductors, the flexible ionic lattice structure results in large polar fluctuations and consequently high dielectric constants,\cite{brivio2013structural, almora2015capacitive} facile screening of excess charges, low exciton binding energies\cite{d2014excitons, miyata2015direct}, and intermediate electron-phonon coupling strengths.\cite{soufiani2015polaronic, wright2016electron} These unique material properties admit the emergence of novel collective behavior, such as an array of photoactive and photo-induced phase transitions.\cite{halder2015exploring,matsuishi2004optical,bischak2019liquid, lin2018thermochromic}

A number of models have been proposed to explain the collected observations of photoinduced phase separation and how it is modulated with changes in composition and large scale morphology.  Initial proposals posited the phase separated state was the equilibrium state at ambient conditions.\cite{brivio2016thermodynamic} In such models, photoexcitation was envisioned to anneal the material towards this equilibrium, however they fail to account for the reversibility of phase separation and the ability to cycle it through alternating light and dark conditions. Other models suggest the driving force for 
phase separation is the band gap reduction associated with halide regions rich in one component. \cite{draguta2017rationalizing} 
While this reduction in band gap is a necessary condition to stabilize halide composition fluctuations, it is insufficient on its own because it does not shift the system from above to below its thermodynamic miscibility critical point. The bias of charge localization towards a lower band gap region would result in a linear change to the local composition that is proportional to the reduction in band gap energy.  If it were the only stabilizing factor for phase separation, the stabilized regions would be made up of compositions that would depend on the initial bulk material composition, which is inconsistent with observations.\cite{hoke2015reversible} 
The emergence of phase separation requires raising of the miscibility critical point, and a mechanism to produce emergent bistability in the free energy, a highly nonlinear response that pins the local composition to a specific value for a given temperature.

The reversible, nonlinear changes accompanying photoexcitation have been suggested previously to proceed through the formation of polarons, which can induce excessive strain energy, destabilizing solid solutions.\cite{bischak2017origin}  Polaronic effects in the lead-halides have been invoked broadly to rationalize the long diffusion lengths of photogenerated charge carriers and the temperature dependence of their mobility, as well as their low bimolecular radiative recombination rates.\cite{zhu2015charge,miyata2017large,neukirch2016polaron,frost2017calculating, miyata2017large} A thermodynamically consistent framework for changes in solid solution phase separation with polaron formation was initially proposed by 
us in
Bischak et al,\cite{bischak2017origin} and a generalization incorporating effects from nanostructured domains was proposed by Wang et al.\cite{wang2019suppressed} This framework clarifies most of the existing observations of phase separation in the lead halides, and is the specific focus of this perspective.
Before focusing on the details of that framework in the next section, we first summarize existing experimental observations, focusing not on completeness but on those experimental facts that have been generally accepted and reproduced. In the subsequent section, 
we review the detailed mechanism of how polarons induce a self-limiting form of phase separation, and following that, we review a model of the dynamics implied by such thermodynamic driving forces. We conclude this perspective by summarizing the relation of the model to the observations and by noting the remaining open questions and challenges. 

\section*{Overview of experimental observations}

\begin{figure}[b]
\begin{center}
\includegraphics[width=8.5cm]{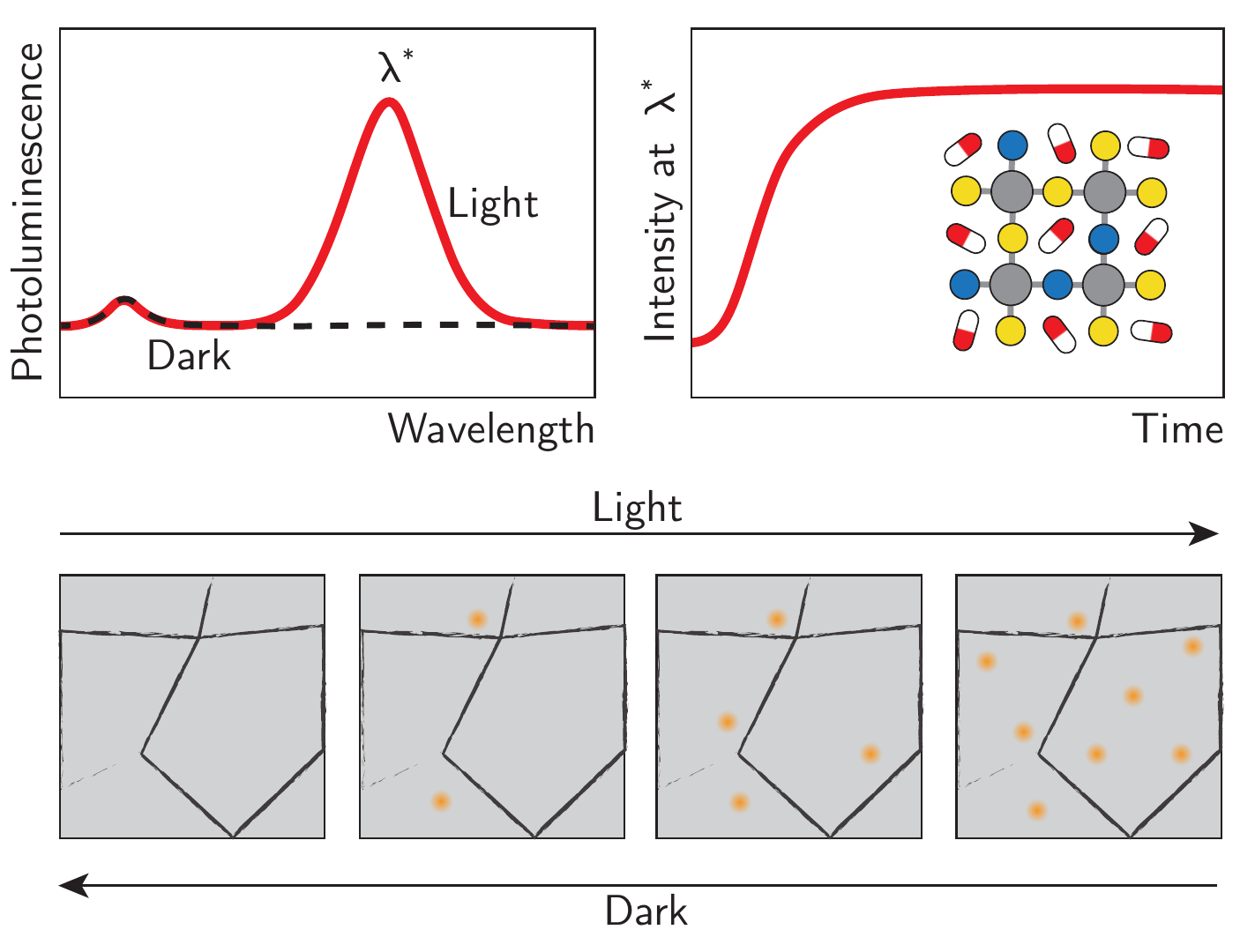}
\caption{Schematic photoluminescence spectra, time dependence and structural changes accompanying photoinduced phase separation. The top panels illustrate the steady state photoluminescence spectra in the dark and under light soaking, as well as how the red shifted peak grows in time. Inset illustrates a cartoon of the perovskite lattice. The bottom panels are representative illustrations of photoluminescence or cathodoluminescence imaging of thin films during photoinduced phase separation.\cite{bischak2017origin,bischak2018tunable} Within individual grains (grey), self-limiting regions enriched in one halide form (yellow). The number of such regions increase upon continued light exposure, but reversibly dissolve in the dark. Adapted from Fig. 1 in Ref.~\onlinecite{bischak2017origin}.}
\label{Fi:1}
\end{center} 
\end{figure}

The basic phenomenology of photoinduced phase separation in the lead halides is illustrated in Fig.~\ref{Fi:1}. As first observed by Hoke et. al.\cite{hoke2015reversible}, an initially well mixed solid solution of MAPb(Br$_x$I$_{1-x})_3$, with $x$ between 0.1 and 0.9 will have a characteristic photoluminescence peak between the extremal values of a pure Br and pure I lattice, reflecting the composition dependent optical gap of the material. Upon continued exposure to light at or above this energy, the photoluminescence peak red-shifts and increases in intensity. At steady state, the location of the peak is coincident with a lattice initially prepared near $x=0.2$, independent of the initial composition. The independence from initial composition indicates the nonlinearity of the effect of illumination. The time dependent rise in the new photoluminescence peak is sigmoidal, with a characteristic timescale typically on the order of minutes. Importantly, nearly all observations of this process report it to be reversible.\cite{hoke2015reversible,bischak2017origin,bischak2018tunable,nandi2018temperature} Provided sufficient time without light exposure, the photoluminescence peak returns to its initial energy. This reversibility indicates that the initial solid solution is stable under low light conditions, but that stability is altered under illumination. 

Accompanying the formation of the low energy luminescent state are dramatic changes to the lattice. Studying thin films, Hoke et al\cite{hoke2015reversible}  reported a splitting of x-ray diffraction peaks under light soaking, and suggested that halides separated into two distinct crystalline phases. Similar x-ray studies followed, confirming that initial observation and indicated the potential for coherency strain, the strain due to lattice mismatch, between regions of different composition.\cite{lehmann2019phase,fang2016photoluminescence,byun2017light} Early work by Bischak et al., also on thin films of MAPb(Br$_x$I$_{1-x})_3$, employed cathodoluminescence imaging and found that small iodine rich aggregates form throughout the film.\cite{bischak2017origin} Rather than growing in size or coalescing, the density of aggregates was found to increase over time until the number reached a steady state value. See Fig.~\ref{Fi:1} for an illustration. At longer times, those aggregates where observed to diffuse to grain boundaries, presumably to reduce the coherency strain previously noted. Subsequent study on single crystals observed that the initial location of iodine rich aggregates was stochastic in repeated measurements.\cite{bischak2018tunable} These observations demonstrated that the formation of iodide rich domains was intrinsic to the material and not dependent on large scale morphology like the presence of grain boundaries. Additional work employing fluorescence imaging in single crystals and photoluminescence imaging in thin films have confirmed these studies.\cite{dequilettes2016photo, mao2019visualizing,wang2019suppressed} In all direct observations, there is no indication of macroscopic phase separation, although diffraction limitations in photoluminescence and carrier migration lengths in cathodoluminescence each preclude direct quantification of domain size. 

The direct imaging of phase separation demonstrated the facile means by which ions can typically move throughout the perovskite lattice. These observations are consistent with AC conductivity measurements that have long demonstrated that purely inorganic perovskites, like CsPbBr$_3$ or CsPbCl$_3$, are efficient halide conductors.\cite{mizusaki1983ionic} Mobile halide ions are also implicated in  observations of hysteresis in cyclic voltammetry studies in hybrid perovskites with organic cations such as MA and FA.\cite{pellet2014mixed, tress2015understanding}  Single crystalline perovskite nanowire heterojunctions made from CsPbBr$_3$ or CsPbCl$_3$ have been studied with wide-field and confocal photoluminescence, allowing for the visualization of halide anion inter-diffusion in the absence of light soaking where the homogeneous solid solution is stable.\cite{lai2018intrinsic} The combination of direct imaging and simple geometry allowed for the accurate measurement of anionic lattice diffusivities, and indicated that halide diffusion was vacancy mediated. Activation energies extracted from those halide diffusion measurements in the dark are consistent with activation barriers for phase separation,\cite{hoke2015reversible} suggesting that phase separation is limited by diffusion and that recent observations of light enhanced conduction\cite{kim2018large} does not seem to play an important role in photoinduced phase separation.

While photoinduced phase separation has been observed in a variety of different morphologies produced under disparate synthesis conditions,  differing morphologies appear to change the rate of phase separation.\cite{zhang2017photoluminescence} By directly modulating halide vacancy levels through the exposure of excess halides, halide mobilities can be decreased and subsequently phase separation effects have been shown to become less prominent over experimental timescales.\cite{yoon2017shift} Similarly, photoinduced phase separation can be partially suppressed by making grains larger, \cite{hu2016stabilized} passivating surfaces with ligands,\cite{xiao2017mixed} or by employing small nanocrystals or films with high quantum yields.\cite{sutter2015high} Indications are thus that the stability of photoinduced phase separation is not strongly influenced by morphology but that the kinetics associated with halide inter-diffusion can be dramatically altered depending on the halide vacancy concentrations resulting from that morphology. 

The dependence of phase separation on composition has received significant attention as a potential route to mitigating its detrimental effects on device performance. For mixed halide perovskites with MA cations, mixtures of both Br/I and Cl/Br have  been found to undergo photoinduced phase separation across broad ranges of compositions.\cite{yoon2016tracking,talbert2017bromine,bischak2017origin,byun2017light}
Under low light conditions, mixtures of Cl and I are not stable under ambient conditions and phase separate into macroscopic domains, indicating the presence of an equilibrium miscibility transition temperature above 300 K for these most size-dissimilar halides.\cite{ralaiarisoa2017correlation} 
While the band structures and mechanical properties of lead halide perovskites are largely insensitive to changes of cation (MA, FA or Cs), there has been much study into the effect of mixing different cations on mitigating photoinduced phase separation. Studies on Cs based perovskites are somewhat mixed, with indications in thin films that they are more stable to light-induced phase separation over some ranges of Br/I ratios $x_\mathrm{Br}<0.33$, though it can still occur for $x_\mathrm{Br}>0.33$.\cite{beal2016cesium,bush2018compositional} Cation replacement of MA with Cs in single crystals has been shown to  stabilize the mixed halide state with only transient I-rich clustering being observed and no stable phase separation.\cite{bischak2018tunable} Resilience to photoinduced phase separation has been demonstrated with mixed cation perovskites including mixtures of FA and Cs.\cite{mcmeekin2016mixed}  Finally, replacing Pb with Sn has also been shown to suppress photoinduced phase separation.\cite{yang2016stabilized} Most of these observations show that more polar lattices are more susceptible to photoinduced phase separation.  In most of these studies, however, changes to the underlying crystal structure stability and defect concentration are not considered.

The dependence of photoinduced phase separation on the thermodynamic state of the material is of specific interest, as such information provides insight into the underlying driving forces and universalities of the transition. The temperature dependence has been studied in both Cs and MA based materials. In MA based films, photoinduced phase separation has been found to occur throughout temperatures ranging between 200 K and 300 K.\cite{hoke2015reversible}  Notably by measuring the initial growth rate of the new photoluminescence peak as a function of temperature, the activation energy for phase separation could be measured. It was found to be similar to the previously measured halide conductivity activation energies, indicating that phase separation itself is a diffusive process. A later study found that below 190 K, phase separation did not occur during 4 hours of illumination, indicating either the slow down of halide diffusion or the role of different structural phases that become stable at low temperature in mitigating phase separation.\cite{nandi2018temperature} That study also found that phase separation ceased at temperatures above 325 K, indicating the location of the light induced miscibility temperature to be just above ambient conditions.  The temperature dependence has also been measured in Cs based crystals, between 150 K and 300 K, where for compositions of $x_\mathrm{Br}>0.4$ phase separation was consistently observed.\cite{wang2019suppressed} As in the MA films, lower temperatures required significantly longer times for phase separation to equilibrate. Very high pressures, greater than 0.9 GPa, have been found to suppress photoinduced phase separation in MA based films.\cite{jaffe2016high}  This threshold pressure corresponds to a change in stability from the normal cubic lattice structure to a so-called $\beta$-phase where neighboring lead-halide octahedra are tilted on average away from 180$^\mathrm{o}$.  The sharp thresholds in pressure, temperature and composition observed are thus consistent with canonical phase behavior, though experimentally determining the phase boundaries with high accuracy is complicated by slow, diffusive kinetics.  The combination of the high dimensional space of thermodynamic parameters specifying the state of the material and these slow kinetics have precluded a systematic experimental construction of the phase diagram under illumination. 

Provided a morphology and composition where phase separation can occur, its extent and characteristic time dependence depends on the illumination intensity and light soaking protocol.
Under constant, continuous wave illumination, the steady state photoluminescence intensity has been shown to be proportional to the carrier concentration, at low power.\cite{bischak2017origin} At high power, the photoluminescence intensity has been shown to saturate, together with the carrier concentration.\cite{draguta2017rationalizing} The growth of photoluminescence intensity reflects a growing number of aggregates formed, as has been visualized with photoluminescence and cathodoluminescence imaging, whose number grow with the steady state carrier concentration before saturating.  Most studies have shown that the time to steady state photoluminescence also depends on the illumination intensity, and Bischak et al demonstrated that it scales linearly with the inverse carrier density for small densities.\cite{bischak2017origin} Analogously, the rate of cluster formation, as visualized with cathodoluminescence imaging, was found to be independent of carrier concentration, reflecting that each cluster forms independently. 

Indications that it is the steady-state concentration of charge carriers that dictates phase separation are supported by varying exposure time studies and observations of phase separation from direct current injection. Specifically, Yang et al. manipulated the charge carrier concentration as a function of time by varying the repetition rate of illumination at constant average power in a pulsed laser experiment.\cite{yang2016light} They observed a threshold frequency of 500 Hz below which no shift of the photoluminescence peak was observed, but above which phase separation occurred. It has been observed that internal gradients in carrier concentrations produced by asymmetric illumination can affect phase separation.\cite{barker2017defect} A definitive measurement that points to the particular role of charge carriers was carried out by Braly et al\cite{braly2017current} and separately by Duong et al,\cite{duong2017light} both of whom demonstrated that running a steady-state current through the film can also induce phase separation when the charge injection was equivalent to that generated by 1 sun illumination. As with transient illumination, transient scanning electron beam experiments do not induce phase separation below a threshold dose.\cite{bischak2017origin} Taken together, these observations indicate that the extent of phase separation depends on the steady-state concentration of charge carriers and suggest that at low, e.g., 1 sun equivalent, intensities each localization of a charge carrier in a clustered region occurs independently. 

This array of experimental observations points to the specific role of excess charge carriers, generated by light or injected directly, in altering the reversible miscibility of a mixed halide solid solution in the lead halide perovskites. Changes in initial temperature, pressure and composition, can have large effects on the extent and appearance of the photoinduced phase separation, while changes in morphology seemingly most affect its kinetics. It would seem  that a natural framework for  rationalizing the driving force for photoinduced phase separation would be to consider how excess charges can modulate an underlying miscibility phase transition. With such a mechanism, the dependence on thermodynamic conditions could be rationalized together with observation of aggregates that are self-limiting. Provided a consistent thermodynamic understanding of photoinduced phase separation, an understanding of the stochastic dynamics that form and dissolve aggregates could then follow. In the next two sections we lay out a specific mechanism for rationalizing the driving force for photoinduced phase separation and its subsequent implications for the associated phase separation kinetics. This framework was originally presented along with experimental observations  in Ref.~\onlinecite{bischak2017origin}, was refined in Ref.~\onlinecite{bischak2018tunable}, and is consolidated in detail in this perspective. 

\section*{Molecular framework for photoinduced phase separation}
In order to develop a microscopic understanding of photoinduced phase separation in lead halide perovskites, we must be able to distinguish the underlying thermodynamic driving forces that would result in phase separation stability, from dynamical effects associated with a hierarchy of disparate timescales in photo-active materials. We therefore first present a framework for considering the thermodynamics of photodinduced phase separation and then present the dynamical considerations afterwards. The thermodynamic perspective we present is based on a generalization of standard solid state solution phase transition theory, adding in a source of strain from photogenerated localized charge carriers. The dynamical perspective is simplified by assuming that halide motion is vacancy-mediated and is the rate limiting step to phase separation, as it occurs orders of magnitude slower than charge carrier and lattice relaxation.  The multi-scale nature of this process requires that novel numerical tools be invented and used to test the simplifying assumptions of the theory. These we also discuss below.

\begin{figure}[t]
\begin{center}
\includegraphics[width=8.5cm]{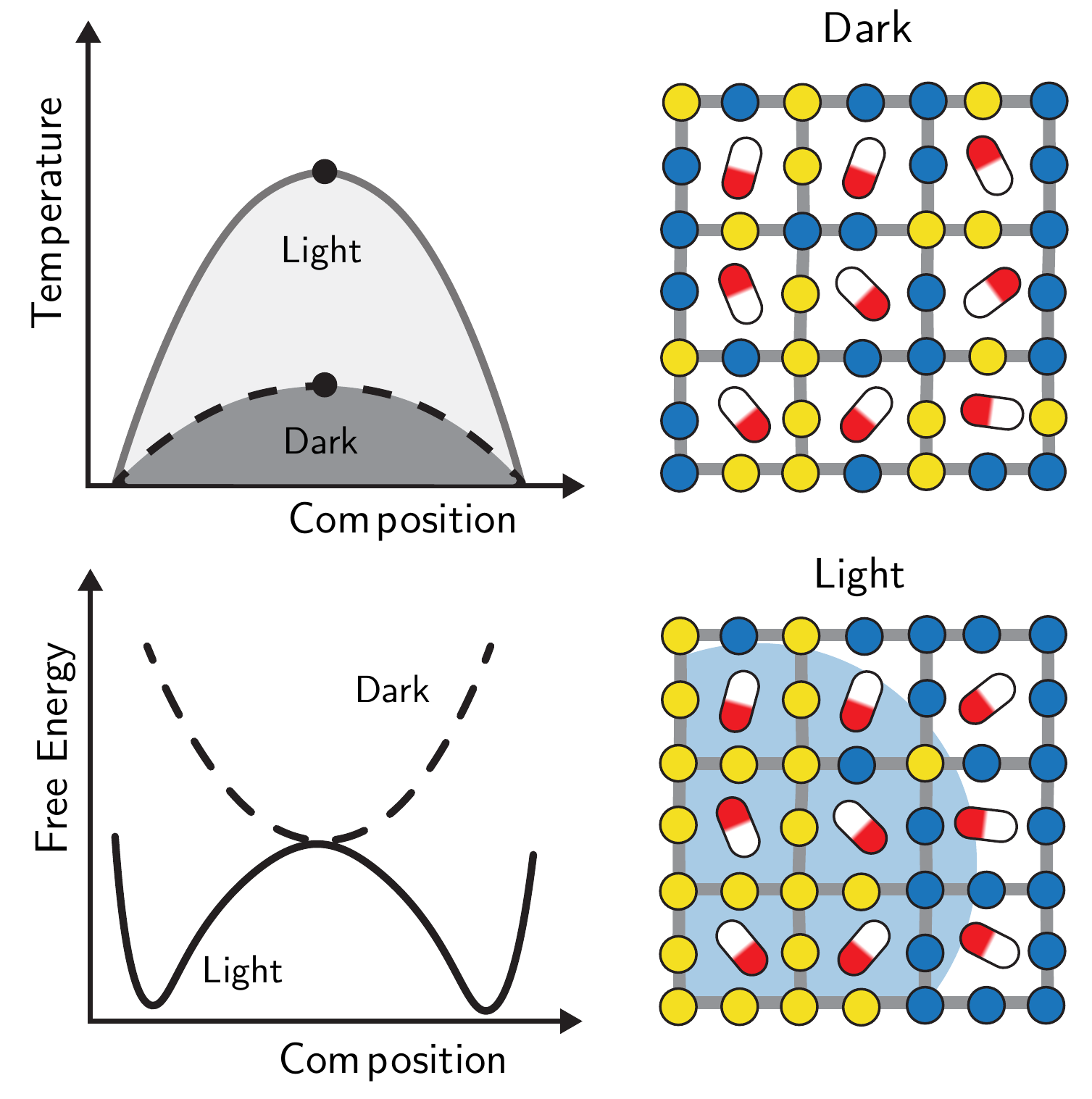}
\caption{Illustration of how the phases diagram (upper-left) change under illumination, due to the change in the free energy surface (lower-left). The right cartoons illustrate the distribution of two halide species (yellow and blue) in the light and dark. Adapted from Fig. 2 in Ref.~\onlinecite{bischak2017origin}.}
\label{Fig:2}
\end{center} 
\end{figure}

\subsection*{Thermodynamics of photoinduced phase separation}

Given the preponderance of experimental observations pointing to the reversibility and robustness of the photoinduced phase separation transition, a description of this process based on modulating an underlying miscibility transition was proposed by Bischak et al.\cite{bischak2017origin} That theory, reviewed here, builds on textbook descriptions of phase separation in solids coupling compositional fluctuations with elastic deformations,\cite{cahn1962spinodal} with an additional field describing the charge density fluctuations generated by free charges. While simplified, the basic physical picture has been verified with explicit molecular simulations as discussed at the end of this section.\cite{bischak2017origin}

In general, the thermodynamics of phase separation of a binary mixture are the result of a balance between the entropy of mixing that tends to favor homogeneous phases and specific energetics that can favor phase separation. We take the local concentration difference, $\phi(\ve{r})=\rho_{X'}(\ve{r})-\rho_{X}(\ve{r})$, between two halide species, $X$ and $X^{'}$, at point $\ve{r}$ in a coarse-grained continuum field to be our local order parameter. For small deviations in composition from $\phi(\ve{r})=0$, the simplest model capable of describing phase separation is the effective Hamiltonian or free energy function,
\begin{equation}
\label{Eq:1}
H_\phi = \int d\ve{r} \, a \phi^2(\ve{r})/2 + u \phi^4(\ve{r})/4+ g| \nabla \phi(\ve{r}) |^2 /2
\end{equation}
where the polynomial expansion of $\phi(\ve{r})$ follows from the expansion of the entropy of mixing and pair interactions for small $\phi(\ve{r})$, and the square gradient term accounts for interface formation, as needed for the conservative matter field considered here.\cite{chaikin2000principles}  For simplicity we work under conditions of zero chemical potential difference between the two halides. 

In principle, all three parameters, $a,u$ and $g$, depend on external conditions. The parameters are determined using a Flory-like model,\cite{chaikin2000principles} with $a=2\kB T-  \Delta H_\mathrm{m}$, where  $\Delta H_\mathrm{m}$ is the enthalpy of mixing at constant volume evaluated at $\phi=0$ in the cubic perovskite phase and $\kB T$ is Boltzmann's constant times temperature, $u=4\kB T/3$, and $g$ is proportional to the surface tension between $X$ and $X^{'}$ rich domains. More generally, the average composition in the phase separated state and the local compressibility of compositional fluctuations above the miscibility temperature can be parameterized in terms of $a$ and $u$. Such a parameterization could be used to differentiate the equilibrium phase diagrams of Cs and MA based lattices, as their miscibility gaps are reported to be different.\cite{bechtel2018first}
Minimization of this effective Hamiltonian results in a phase diagram that at low temperature supports phase separation, at high temperature predicts that the mixture is miscible, and at an intermediate temperature predicts a miscibility temperature that delineates these two regions. 
See for example Fig.\,~\ref{Fig:2}.

Since there are large differences in the lattice constants between any two pure halide phases in cubic APbX$_3$ structures, the effects of coherency strains generated between two phase separated lattices cannot be ignored. Provided that there are no large scale morphological transitions accompanying phase separation and that the cubic symmetry of the perovskite lattice does not change upon changing composition, we can restrict the effects of strain to those derivable from linear elasticity. Moreover, we will employ isotropic elasticity theory for simplicity of notation, though our results are insensitive to those generalized to the cubic symmetry of the perovskite lattice with the correct average elastic constant. Under these simplifying assumptions, the contribution to the system from elastic deformations is given by the contribution to the Hamiltonian, 
\begin{equation}
\label{Eq:2}
H_\Sigma = \int d\ve{r} \, K \frac{\Sigma^2(\ve{r})}{2(1-\nu) }
\end{equation}
where $\Sigma(\ve{r})$ denotes the local strain field, $K$ is the bulk modulus in the mixed phase and $\nu$ is Poisson's ratio. For variations in composition, and assuming the lattice is always in instantaneous equilibrium with local compositional fluctuations, the resultant strain due to the lattice mismatch is given to lowest order by
\begin{equation}
\label{Eq:3}
\Sigma(\ve{r}) =  2\eta \phi(\ve{r})
\end{equation}
where $\eta = d \ln \ell /d \phi$ is the change in the lattice constant $\ell$ with composition, evaluated in the mixed phase. Higher order effects, like the change of the bulk modulus with composition, or the violation of Vegard's law are neglected here.\cite{atourki2017impact, weber2016phase} 

As first discussed by Cahn in the context of metal alloys,\cite{cahn1962spinodal} the excess strain due to compositional fluctuations stabilizes the well mixed solid solution by lowering the critical miscibility temperature. We refer to this transition temperature as the miscibility temperature for the lattice in the dark, $T_c^{(\mathrm{d})}$. This temperature can be computed within mean field theory as,
\begin{equation}
\label{Eq:Tcd}
T_c^{(\mathrm{d})} = \frac{1}{2\kB} \left ( \Delta H_\mathrm{m} - \frac{2\eta^2 K}{1-\nu} \right ),
\end{equation}
which follows from functional minimization of the combined Hamiltonians in Eqs.~\ref{Eq:1}-\ref{Eq:3} with respect to the compositional field. While the enthalpy of mixing at constant volume has not been measured experimentally, calculations have been performed at various levels of accuracy.  Typical values for APbX$_3$ for mixtures of Br and I are between 2-4 kcal/mol.\cite{bischak2017origin,bechtel2018first} The bulk modulus, $K$, is known experimentally for a variety of compositions and is typically close to 15 GPa.\cite{rakita2015mechanical} The change in the lattice constants with composition are also well known, for example for MAPbX$_3$ mixtures of Br and I, $\eta=0.08$ and for CsPbX$_3$ $\eta=$0.06.\cite{lehmann2019phase,mashiyama1998disordered} Using Eq.~\ref{Eq:Tcd}, these material properties imply critical miscibility temperatures in the dark all below ambient conditions, in agreement with observations of stable well mixed solid solutions under those conditions. The contributions from the strain relaxation in all cases are large, with $\eta^2 K /(1-\nu) \kB$ being on the order of 800 K, indicating the large susceptibility of the miscibility transition to strain. 

Under illumination, excess charge carriers change this traditional picture of phase separation in two ways. First, charges are attracted to regions rich in the component with the lower band gap. Second, provided charges form polarons, they strain the lattice, which as we have just shown, can greatly affect the stability of the mixed phase. The coupling of the charge to the composition field is straightforward, and computable from the band offset energies in two limiting pure halide lattices. Accounting for the strain of the lattice is more complicated. 
Polar coupling to the optical phonons is known to contribute to electron-phonon interaction in the perovskites,\cite{wright2016electron} and there are observations of acoustic scattering as well.\cite{mante2017electron,guo2017polar} 
Moreover, the anharmonicity of the lattice\cite{brivio2015lattice, marronnier2018anharmonicity,yaffe2017local,beecher2016direct} calls into question simplified microscopic models of the electron phonon coupling, which assume a harmonic lattice.
Indeed, more detailed molecular simulations with fewer underlying assumptions illustrate that electron phonon coupling in the perovskites is not simply 
describable as limiting cases of either Peierls, Frohlich, or Holstein coupling.\cite{mayers2018lattice}

We consider the coupling of the charge to the lattice from the perspective of continuum elasticity and use a spatially local first order perturbation theory for electron-phonon interactions.\cite{whitfield1976interaction} This affords a simple description of the excess charge density and its local effect on the strain and composition fields, and should be thought of as a Gaussian fluctuation approximation of the lattice rather than an assumption of it being harmonic. However, as discussed at the end of this section, our previous work 
using detailed molecular models,\cite{bischak2017origin,bischak2018tunable} without assumptions as to the particular form of the electron-phonon coupling validate this basic approach. 
 To make the calculation tractable, we make a semiclassical approximation to decouple the electronic and lattice motions, and, as before, assume the lattice deformations are constantly in equilibrium with the local composition. Specifically, we add the lowest order coupling between excess charge density amplitude, $\psi$, and the strain and composition fields,
\begin{equation}
\label{Eq:5}
H_\psi = \int d\ve{r} \,  \frac{\hbar^2}{2 m} |\nabla \psi(\ve{r})|^2 - \epsilon \psi(\ve{r}) \phi(\ve{r}) - \alpha \psi(\ve{r}) \Sigma(\ve{r})
\end{equation}
where $\epsilon$ is the energetic bias of an excess charge to reside in different $\phi$ regions, $\alpha$ is the electron-phonon coupling constant, and $m$ is the effective mass of the excess charge and $\hbar$ is Planck's constant divided by $2\pi$. 

Assuming that the charge is localized to a lengthscale, $R_\mathrm{p}$, we can integrate out its effect on the compositional fluctuations in the vicinity of that localized charge.  Within a mean field approximation and working along the symmetry line where $\epsilon =0$, we can compute the critical miscibility temperature under illumination, $T_c^{(\mathrm{l})}$, due to the excess charge. The lowest order correction to Eq. \ref{Eq:Tcd} appears at the level of $\alpha^2$ and is given by,
\begin{equation}
\label{Eq:Tcl}
T_c^{(\mathrm{l})} =T_c^{(\mathrm{d})}  + \frac{2 m \eta^2 \alpha^2 R_\mathrm{p}^2}{h^2 \kB} ,
\end{equation}
which demonstrates that the miscibility temperature increases due to the formation of the polaron by an amount dependent on the electron phonon coupling, $\alpha$, the size of the polaron, $R_\mathrm{p}$, and the lattice mismatch, $\eta$.  Under these assumptions, the stabilization of the phase separated state is due to the strain generated by the charge-lattice coupling, which reverses the reduction of the miscibility temperature upon strain relaxation in the dark.
Typical values of $m=0.15 m_\mathrm{e}$,\cite{herz2017charge} $\alpha=4$ eV,\cite{mante2017electron} and $R_\mathrm{p}=2.5$ nm,\cite{frost2017calculating} result in a predicted enhancement of the miscibility temperature of $T_c^{(\mathrm{l})} - T_c^{(\mathrm{d})}$ = 700 K for MA based perovskites, and 350 K for Cs based materials. This predicted  stability of the demixed phase is consistent with experimental observations of phase separation at room temperature under illumination and also in observing that the stability increases with electron-phonon coupling.\cite{bischak2018tunable} 

Unlike traditional equilibrium miscibility, the light-induced stabilization of the demixed phase is confined to the region of the polaron, as only in its presence is there a source of strain and therefore a driving force of phase separation. This explains why phase separation is observed to be self-limiting.\cite{bischak2018tunable} Photoluminescence measurements done at different powers have inferred a domain size to be 3 nm, assuming spherical clusters.\cite{bischak2017origin} 
The extent of the predicted stabilization from this simplified mean field theory is likely an overestimate, given observations of an upper-limit to photoinduced phase separation of 325 K in MA based films.\cite{nandi2018temperature} Corrections for fluctuations, surface tension from the finite size aggregates, and strain modulation\cite{wang2019suppressed} all tend to destabilize phase separation but are not accounted for in this simple theory.  Nevertheless these arguments provide a useful framework for rationalizing experimental trends, and these corrections could subsequently be computed to achieve tighter agreement of the temperature dependence.

\begin{figure}
\begin{center}
\includegraphics[width=8.5cm]{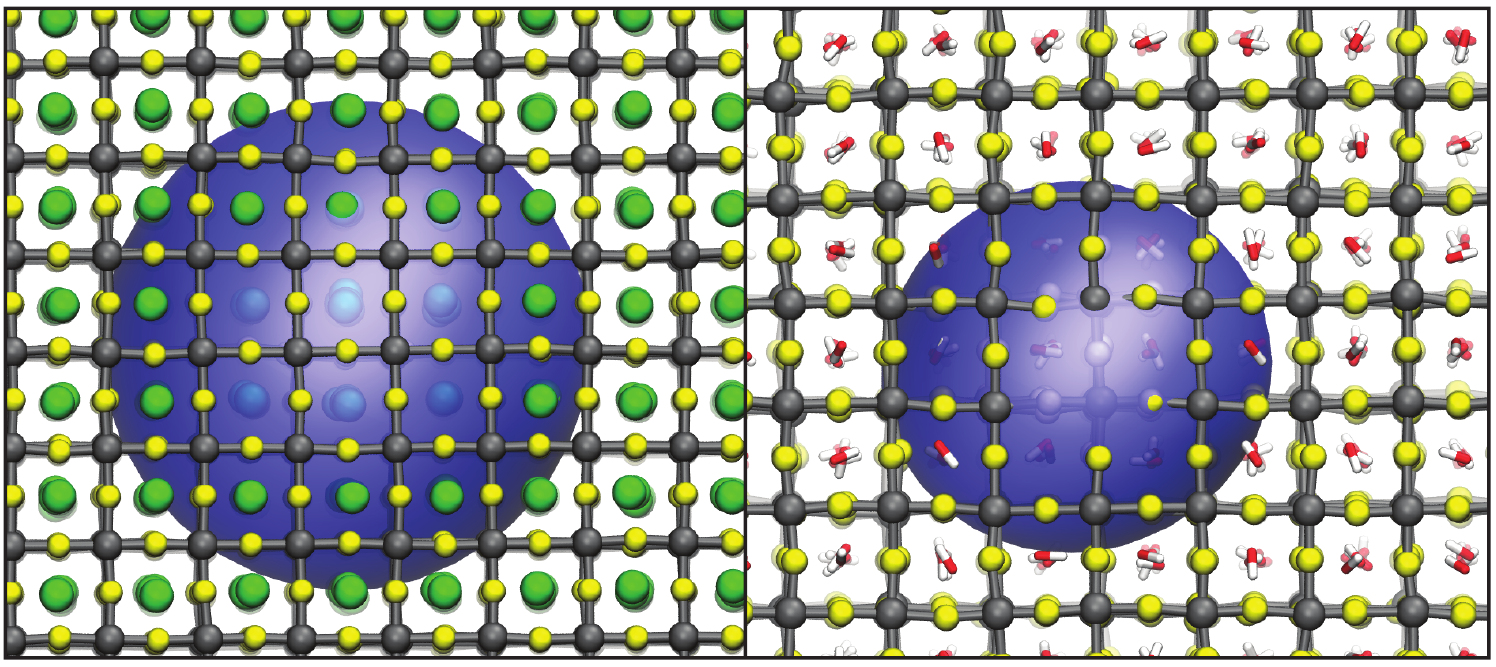}
\caption{Characteristic snapshots of an excess electron in an CsPbI$_3$ (left) and MaPbI$_3$ (right) lattice. The charge is depicted as an isodensity surface in blue from path integral molecular dynamics simulations. Adapted from Fig. 4 in Ref.~\onlinecite{bischak2018tunable}.}
\label{Fi:2a}
\end{center} 
\end{figure}

\begin{figure*}[t]
\begin{center}
\includegraphics[width=13cm]{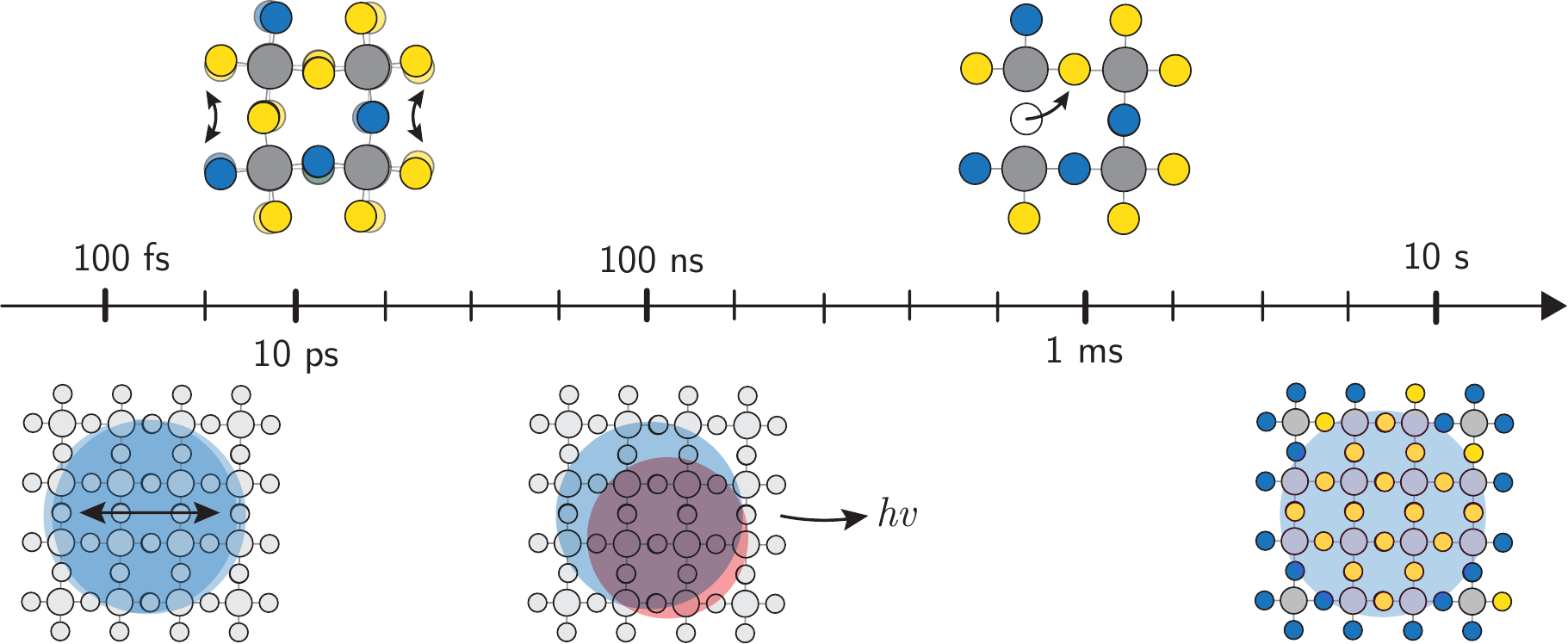}
\caption{Illustration of the broad range of timescales involved with photoinduced phase separation. On the fast timescales, charges move through the lattice, and the lattice vibrates. On intermediate timescales, charges undergo radiative recombination and halides diffuse. Finally, on the longest timescale, an average localized charge density can stabilize locally phase separated regions rich in one halide component.
}
\label{Fi:3}
\end{center} 
\end{figure*}

The above theory makes a number of simplifying assumptions to arrive at a compact expression for the modulation of the miscibility temperature by the formation of a polaron. To test these assumptions computationally and go beyond them, requires a way to represent an extended lattice, and a facile means to sample over different halide compositions and thermal fluctuations to compute the subsequent free energy for differing compositions. While simulations under these conditions are routine with molecular simulation, the photoinduced phase transition additionally requires a means to represent the effect of an excess charge on this free energy. While in principle \emph{ab initio} based molecular dynamics simulations together with enhanced sampling methods could be used, such would be prohibitively computationally expensive for the system sizes required to not artificially localize the excess charge. An alternative method developed in Ref.~\onlinecite{bischak2017origin} used an empirical forcefield for the MAPbX$_3$ lattice,\cite{mattoni2016modeling} together with a path integral description of the excess charge within an effective mass approximation\cite{shumway2004quantum} and a set of empirical pseudo-potentials.\cite{kuharski1988molecular,schnitker1987electron} Such a classical simulation depicted in Fig.~\ref{Fi:2a}, albeit in an extended phase space, afforded the ability to use robust sampling tools and a Grand Canonical ensemble to compute the free energy as a function of composition with and without the excess charge. This calculation confirmed the basic mechanism of polaron stabilized phase separation\cite{bischak2017origin}, relaxing the assumption of specific electron-phonon coupling and incorporating directly an anharmonic lattice. Further simulations demonstrated how cation exchange with Cs could decrease the driving force at ambient conditions by reducing the electron-phonon coupling.\cite{bischak2018tunable}

\subsection*{Kinetics of photoinduced phase separation}
During photoinduced phase separation, motions on vastly different timescales become coupled. 
The lead halide perovskites have low cohesive energies allowing for large scale motions of the lattice, in addition to the expected facile motion of photoexcitations and their dissociated charge carriers for a functioning photovoltaic. 
The basic processes, as well as their characteristic timescales that span 15 orders of magnitude at room temperature, are summarized in Fig.~\ref{Fi:3}. To relate processes over this substantial range of scales, the approach proposed in Bischak et al,\cite{bischak2017origin} was to identify the relevant time scale separations that enable a simplified picture for aggregation to emerge. Below, we review the current understanding of the timescales for various related processes as well as the derivable implications of the preceding thermodynamics section that these dynamics imply.

The fastest process relevant to photoinduced phase separation is the diffusion of photo-excited charge carriers throughout the lattice. Charge diffusion has been studied by a variety of techniques and for lead halide perovskites at ambient conditions, estimates typically place diffusion constants between 0.1-1 cm$^2$/s.\cite{brenner2015mobilities,herz2017charge,delor2019imaging} This implies a characteristic timescale to move between two units cells on the order of 0.01 ps. The precise value depends weakly on the composition, and morphology, and is contained within an order of magnitude for all of Cs-based and MA-based single crystals and thin films. These fairly large diffusivities reflect the small polaron 
binding energies that are not strong enough to cause the charge to self trap. 
At typical illumination intensities, the charge carrier lifetimes are around 100 ns, with radiative recombination acting as the largest loss channel.\cite{herz2016charge,de2015impact,wehrenfennig2014high} The origin of this low bimolecular recombination rate is currently debated.\cite{davies2018bimolecular} On similarly fast timescales are local lattice dynamics. The longitudinal optical modes of lead halides have frequencies between 50 cm$^{-1}$ and 100 cm$^{-1}$,\cite{sendner2016optical,miyata2017lead} with characteristic timescales on the order of picoseconds. Slower octahedral tilting motions occur on timescales of 10 ps.\cite{guo2017interplay} Given the more than 10 orders of magnitude separating these fast motions and the seconds-to-minutes scale of phase separation, it is clear that these degrees of freedom can be assumed to stay in local equilibrium. 

Halide diffusion and interdiffusion occurs on significantly longer timescales than charge or lattice motion. Halide diffusion is vacancy mediated, and thus depends on both the vacancy concentration and the activation energy to move a vacancy between two lattice sites.  Vacancy concentrations, $\rho_\mathrm{v}$, depend strongly on material synthesis conditions. At room temperature, $\rho_\mathrm{v}$ has been estimated to be as low are 10$^{-9}$ per unit cell in single crystals and as high as 10$^{-4}$ per unit cell in thin films.\cite{walsh2015principles,lai2018intrinsic} Activation energies for halide diffusion have been measured experimentally, and estimated theoretically, and fall within a range of 0.2 - 0.4 eV.\cite{walsh2015principles,lai2018intrinsic} For phase separation, halide interdiffusion is the pertinent transport process. Because halide hopping rates are significantly higher than cation hopping rates, electroneutrality dictates that the self-diffusion constants of each halide in a mixture are the same. The interdiffusion coefficient is given by 
\begin{equation}
\label{Eq:7}
D(\phi)  =\frac{1}{2}\ell^2(\phi) \rho_\mathrm{v}(\phi) k_\mathrm{v}(\phi)
\end{equation}
where $\ell$, $\rho_\mathrm{v}$, and $k_\mathrm{v}$ are the lattice spacing, the vacancy concentration per unit cell, and an effective hopping rate that all depend on the relative concentration of two different halide species.\cite{belova2004analysis,lai2018intrinsic} The dependence of the lattice spacing on composition via $\ell(\phi)$  can be approximated using Vegard's law.\cite{weber2016phase} The vacancy concentration, $\rho_\mathrm{v}(\phi)$, depends on the system morphology and synthesis conditions, in addition to composition through the varying formation free energies.
The effective hopping rate is computable from an average of the hopping rates of each pure lattice component. See Ref.~\onlinecite{belova2004analysis} for a discussion of the theory, and Ref.~\onlinecite{lai2018intrinsic} for the details for Cs-based perovskites. The interdiffusion coefficient for CsPb(Cl$_x$Br$_{1-x}$)$_3$ has been measured in nanowires 
to be 10$^{-12}$ cm$^2$/s, implying that a halide swaps its lattice position with a vacancy once every 1 ms on average, in correspondence with Eq. 7.\cite{whalley2017perspective,lai2018intrinsic}

Under typical conditions, 1 sun illumination and in thin films, phase separation occurs on a timescale of seconds to minutes.\cite{bischak2017origin,hoke2015reversible} Therefore, the characteristic time to form an aggregate of one halide species is much longer than the average lifetime of a charge carrier. Moreover, the diffusion of a charge is about 12 orders of magnitude faster than the diffusion of a halide. Because of this large separation of timescales, the rate limiting step to form an aggregate is the spontaneous fluctuation of the halides creating a region of pure enough halide composition to trap  a charge. Once the charge is trapped it can stabilize the aggregate and ripen it to its equilibrium composition. Within the thermodynamic model above, one charge is sufficient to stabilize an aggregate. After about 100 ns, a trapped charge will undergo radiative recombination, however under constant illumination, another charge, diffusing orders of magnitude more quickly than the halide ions, can become trapped in the aggregate and stabilize it again before the aggregate has the time to dissolve.

Using transition state theory, we can estimate the mean time to form an aggregate. The rate is given by the product of the probability to have a composition fluctuation large enough to trap a charge carrier and the characteristic timescale for halide interdiffusion. Specifically, we assume that a charge localizes in a halide compositional inhomogeneity that forms spontaneously and transiently. The probability to have a specific composition fluctuation in a volume $V$ using the free energy model in the previous section is $p(\phi) = \exp[-\beta a \phi^2 V/2]$. In the presence of the energetic bias towards a specific halide composition due to favorable band energies ($\epsilon$ in Eq.~\ref{Eq:5}), the equilibrium state of a localized charge is defined by $\phi = \epsilon/a$. 
Combining this composition--bias relationship with
 the characteristic attempt frequency for clustering from the halide interdiffusion constant (Eq.~\ref{Eq:7}), the rate for a single aggregate to form is  
\begin{equation}
k_c=\frac{2 D}{\ell^2} e^{-\beta \kappa},
\end{equation}
where $\kappa= \epsilon^2 V/2a$ is the free energy to form a cluster large enough to localize the charge. If we take $\epsilon =0.44 \kB T$, per charge of size $R_\mathrm{p}$, which follows from the difference in the band gap between a pure Br and pure I lattice\cite{noh2013chemical,sadhanala2015blue} and the value of $a$ used to predict the phase diagram for MAPb(Br$_x$I$_{1-x}$)$_3$, the barrier is approximately 0.24 eV.  Using the interdiffusion constant of 10$^{-12}$ cm$^2$/s,\cite{whalley2017perspective} the rate is 0.01 s$^{-1}$, which is in good agreement with that extracted by cathodoluminescence, where single cluster data is available.\cite{bischak2018tunable}  
Synthetic strategies that deplete halide vacancies, like using halide-rich solution conditions, can directly modulate the time to phase separate, as the rate is directly proportional to the vacancy concentration.
Under the assumption that clusters are dilute, so that clusters form independently, the rate to form $N_c$ clusters is $k = k_c N_c$. Given the thermodynamic calculation that illustrates one charge on average is sufficient to stabilize a local phase separated aggregate, this implies that the rate to reach a steady state photoluminescence will be proportional to the steady-state concentration of charges as has been observed previously.\cite{bischak2017origin}

\begin{figure}
\begin{center}
\includegraphics[width=8.5cm]{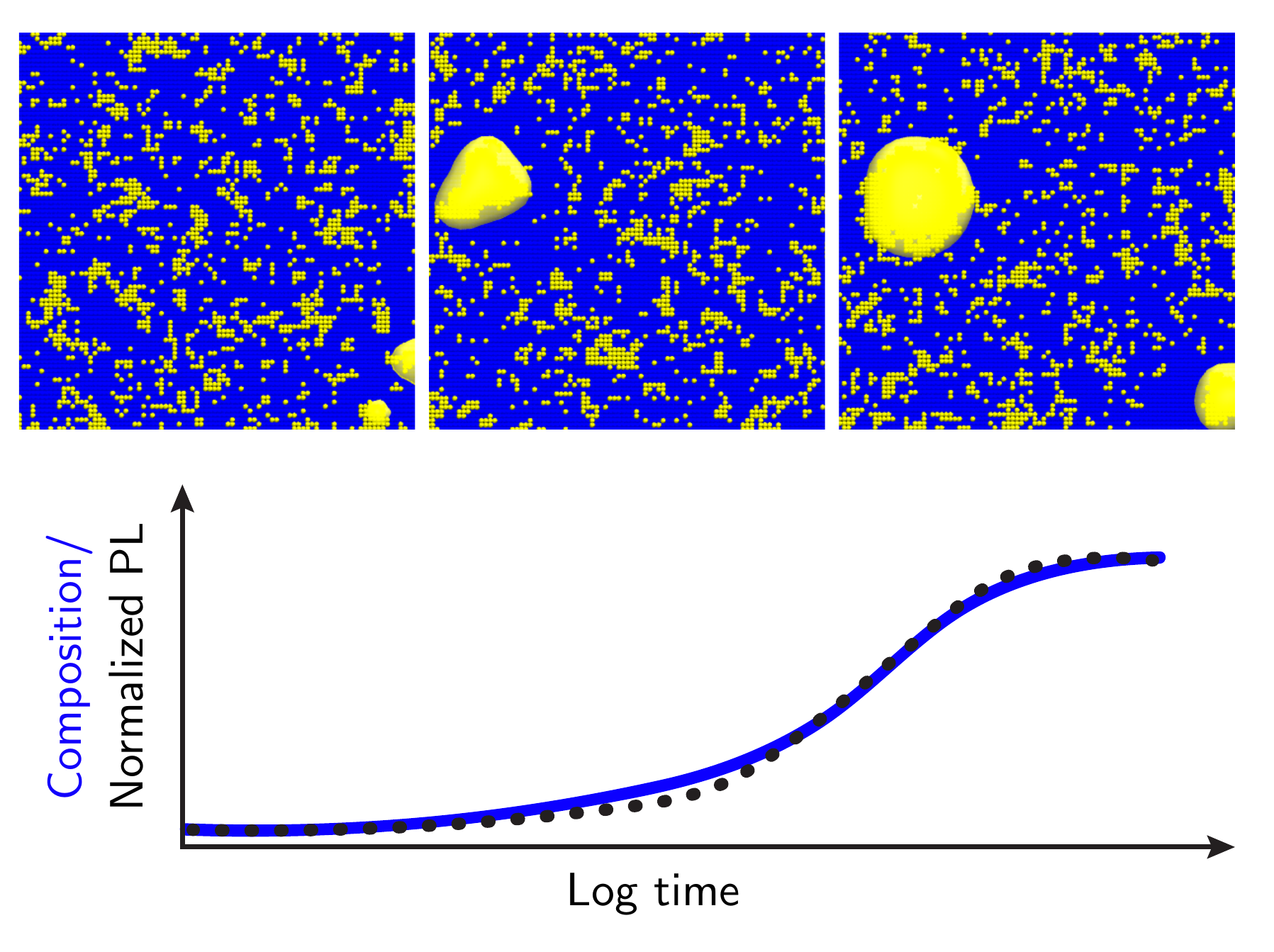}
\caption{Coarse-grained model for photo-induced phase separation. (top) Snapshots of the lattice model used to study photoinduced phase separation taken 30 seconds apart. The blue and yellow lattice sites indicate a locally bromide or iodide rich lattice, respectively. Also shown are yellow isosurfaces for the excess charge densities. The simulations are 2D to represent thin films with periodic boundary conditions and are 2500 nm$^2$ in area. (bottom) Average iodide composition at the excess charge (blue solid curve) and normalized photoluminescence intensity at the characteristic iodide rich optical frequency (black dotted curve) as functions of time. 
Adapted from Fig. 4 in Ref.~\onlinecite{bischak2018tunable}.}
\label{Fi:3a}
\end{center} 
\end{figure}
In order to validate the basic kinetic model, a coarse-grained lattice representation of the system was developed in Refs.~\onlinecite{bischak2017origin,bischak2018tunable}. This simplified model encoded the thermodynamics of Eqs. \ref{Eq:1}-\ref{Eq:5}, with vacancy mediated halide interdiffusion dynamics and explicit charge motion. Specifically, after integrating out the elastic degrees of freedom, an Ising model provided a useful representation of the residual compositional degrees of freedom. It was combined with an additional charge density field, consistent in spatial scale with the results from the path integral molecular dynamics calculations above (Fig.~\ref{Fi:2a}, and that locally renormalized the Ising-coupling strength for lattice sites overlapping it.  Such a reduced representation was necessary in order to span time and length scales of clustering, which are prohibitively large relative to those approachable from the more molecular representations used to established the underlying thermodynamic driving forces for phase separation depicted in Fig. \ref{Fi:3a}.  Explicit parameterizations of the elevation of the local critical temperature in the presence of the charge and a local bias to halide rich regions were both included. A discrete time kinetic Monte Carlo procedure\cite{kawasaki1972phase} was used to update the compositional configuration of the lattice and the location of the charge with characteristic rates of halide diffusion and charge motion taken from experiment.

Such a minimal description of the dynamics of photoinduced phase separation was nevertheless able to reproduce the rise in the emergence and stabilization of isolated I clusters in time-lapse cathodoluminescence imaging, the concomittant photoluminescence intensity versus time as the characteristic time to cluster\cite{bischak2017origin} as well as the steady state spectra as given by the steady state overlap of the charge with the local composition.\cite{bischak2018tunable} Simulations of that model, shown in Fig.~\ref{Fi:3a}, confirmed that the rate limiting step of cluster formation was a transient composition fluctuation, creating an aggregate large enough to trap a charge, with energetic bias $\epsilon$. The localized charge thereafter stabilizes the composition fluctuation by locally placing the lattice  below the critical miscibility temperature. This feedback mechanism is thus unlike either nucleation and growth or spinodal decomposition.  As a charge is only trapped transiently, having a lifetime of only around 100 ns,
the aggregate stabilization dynamics are intermittent, including cycles of different charges being serially trapped,  stabilizing aggregation, and recombining. A given aggregate may persist over many such cycles as charges are replenished during steady-state illumination faster than the aggregate can dissolve. These cyclic dynamics are a consequence of the nonequilibiurm steady-state nature of the phase separation. The subsequent intermittency was explored in Ref. ~\onlinecite{turaeva2018ising} and visualized directly in single crystals using photoluminescence imaging.\cite{bischak2018tunable}

\section*{Concluding remarks}
The photoinduced phase separation of multi-component lead halide perovskites is well documented experimentally as a robust, reversible phenomena. Within the framework presented here, it 
emerges due to the confluence of unique material properties. The soft perovskite lattice admits high vacancy concentrations, and the ionic lead halide bonding couples to photoinduced charges, creating large polarons. When an initially homogeneous solid solution of different halides are present, the formation of a polaron generates sufficient strain to locally render phase separation thermodynamically favorable. High vacancy concentrations result in large interdiffusion coefficients, which renders phase separation observable on experimentally relevant timescales. 

While the framework expanded upon here includes a number of simplifying assumptions that preclude perfect agreement with experimental observations of quantities such as light/polaron-induced critical temperature shifts, these assumptions accommodate the challenging multiscale nature of the photo-induced phenomenon. The model already offers a way to rationalize the majority of the experimentally observed behavior consistently and also consistently describes how the behavior can be modulated by material composition, morphology and thermodynamic state. In the future, the model could be further refined to include corrections for fluctuations, surface tension from finite size aggregates, strain modulation, or high charge carrier concentrations. 
Yet, with the explicit relations provided above, light-stable perovskite based materials  can already be rationally designed for a variety of photophysical applications. Strategies that have and will continue to be pursued will either manipulate the stability of the phase separated state by modulating the critical miscibility temperature under illumination, or reduce the rate at which it is formed, by pursuing materials with lower vacancy concentrations.  

\section*{Acknowledgements}
DTL acknowledges support under the U.S. Department of Energy, Office of Science, Office of Basic Energy Sciences, Materials
Sciences and Engineering Division under Contract No.
DE-AC02-05-CH11231 within the Physical Chemistry of
Inorganic Nanostructures Program (KC3103). 
N.S.G. acknowledges support from the Science and Technology Center on Real-Time Functional Imaging, a National Science Foundation Science and Technology Center (grant DMR 1548924), the Photonics at Thermodynamic Limits Energy Frontier Research Center funded by the US Department of Energy, Office of Science, Office of Basic Energy Sciences (award DE-SC0019140), a David and Lucile Packard Fellowship for Science and Engineering, an Alfred P. Sloan Research Fellowship, and a Camille Dreyfus Teacher-Scholar Award.

\section*{References}

%

\end{document}